\documentclass[sigconf]{acmart}
\AtBeginDocument{%
  \providecommand\BibTeX{{%
    \normalfont B\kern-0.5em{\scshape i\kern-0.25em b}\kern-0.8em\TeX}}}

\copyrightyear{2024}
\acmYear{2024}
\setcopyright{acmlicensed}\acmConference[In2Writing '24]{The Third Workshop on Intelligent and Interactive Writing Assistants}{May 11, 2024}{Honolulu, HI, USA}
\acmBooktitle{The Third Workshop on Intelligent and Interactive Writing Assistants (In2Writing '24), May 11, 2024, Honolulu, HI, USA}
\acmDOI{10.1145/3690712.3690724}
\acmISBN{979-8-4007-1031-5/24/05}

\begin{document}

\title[Ink and Individuality]{Ink and Individuality: Crafting a Personalised Narrative in the Age of LLMs}

\author{Azmine Toushik Wasi}
\authornote{Contact author.}
\orcid{0000-0001-9509-5804}
\affiliation{%
  \institution{Shahjalal University of Science and Technology}
  \city{Sylhet}
  \country{Bangladesh}
  }
  \email{azminetoushik.wasi@gmail.com}
  
\author{Raima Islam}
\orcid{0009-0007-4744-9015}
\affiliation{%
  \institution{BRAC University}
  \city{Dhaka}
  \country{Bangladesh}
}
\email{raima.islam@g.bracu.ac.bd}

\author{Mst Rafia Islam}
\orcid{/0009-0000-2561-3418}
\affiliation{%
  \institution{Independent University}
  \city{Dhaka}
  \country{Bangladesh}
  }
\email{rafiabarsha21@gmail.com}

\renewcommand{\shortauthors}{Wasi et al.}

\begin{abstract}
Individuality and personalization comprise the distinctive characteristics that make each writer unique and influence their words in order to effectively engage readers while conveying authenticity. However, our growing reliance on LLM-based writing assistants risks compromising our creativity and individuality over time. We often overlook the negative impacts of this trend on our creativity and uniqueness, despite the possible consequences.  This study investigates these concerns by performing a brief survey to explore different perspectives and concepts, as well as trying to understand people's viewpoints, in conjunction with past studies in the area. Addressing these issues is essential for improving human-centered aspects of AI models and enhancing writing assistants for personalization and individuality.
\end{abstract}





\maketitle
\section{Introduction}
\begin{figure}[t] 
\centering {\includegraphics[scale=0.15]{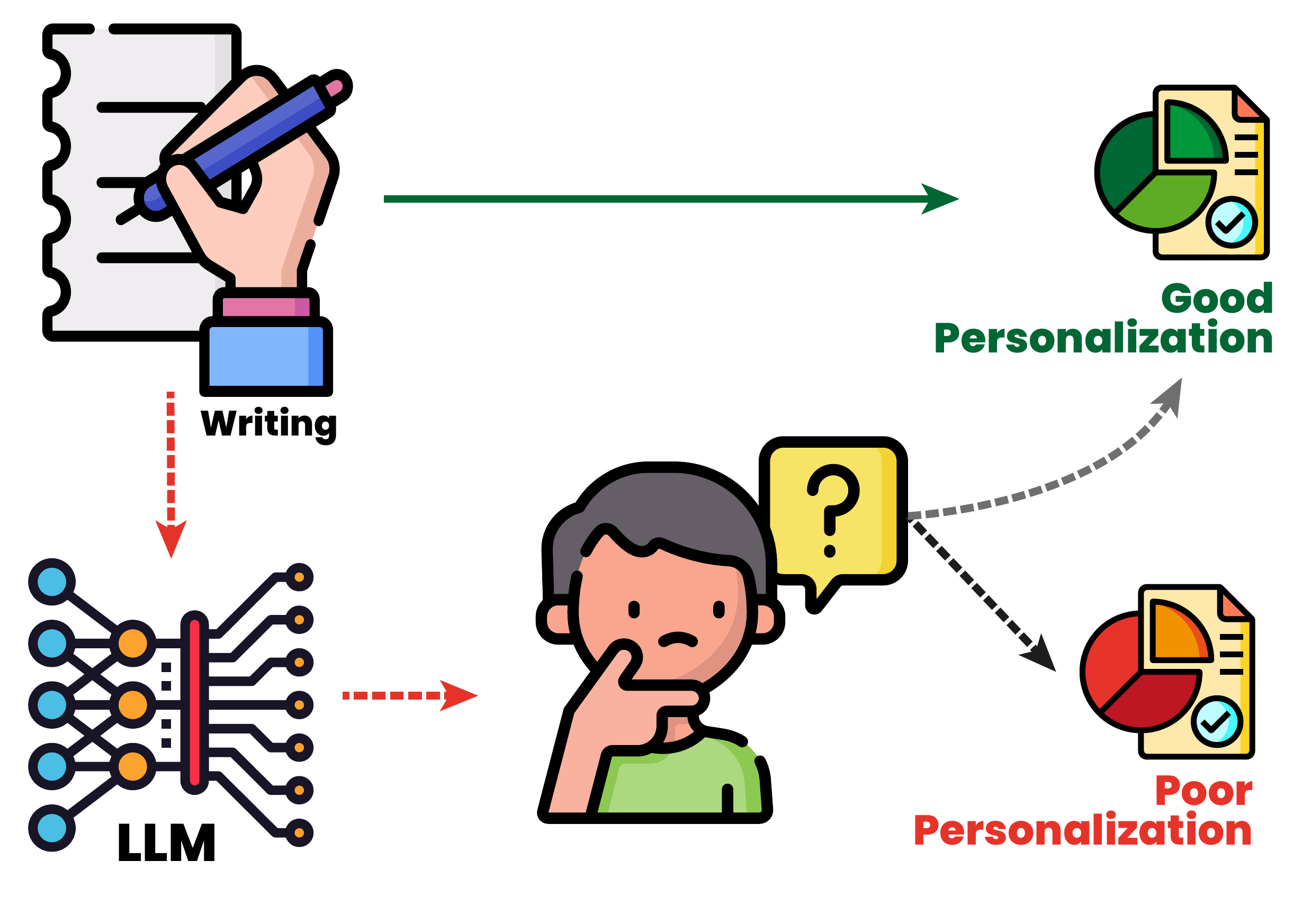}}
\caption{Crafting a Personalised Narrative in the Age of LLMs: \textit{Is it better or worse?}}
\label{fig:WHAT}
\end{figure}

The exceptional performance of large language models (LLMs) across a wide range of applications makes them increasingly well-liked in academia and industry \cite{chang2023survey-1}. Especially with chatbots such as Microsoft Copilot, Gemini, and ChatGPT, which have the transformative capability of producing polished and well-crafted stories, essays and assignments akin to those written by a professional novelist or a writer \cite{10.1145/3491102.3501819, wordcraft}. Not to mention, other writing tools such as Quillbot and Grammarly have made writing a straightforward, time-efficient task now. With recent works such as \cite{kim2023lmcanvas} and \cite{yeh2024ghostwriter}, we are at a junction where personalization is of high significance with LLM usage for writing. This raises a question: Are users concerned about the level of personalization in their writings with the emergence of AI-assisted writing tools? (Figure \ref{fig:WHAT})

In this paper, we explore (i) the importance of personalization and individuality of writing and (ii) personalization in the age of LLMs. In this study of these issues, we try to understand the perspective of individuals through a thoroughly conducted survey, along with previous research work done in this domain. We aim to contribute to the development of personalization in writing with LLMs along with its significance and try to highlight whether individuals are affected by it. It is important to highlight the personal touch of an individual when writing content, for example, crafting personal statement essays for university applications or cover letters for jobs, which are important factors in whether someone gets accepted or rejected. 

\section{PERSONALIZATION AND INDIVIDUALITY
IN WRITING}
Personalization in writing means customizing the content to meet the needs and preferences of the audience, while still maintaining the author's distinct voice and perspective. This includes adjusting the language, tone, and examples to connect with various readers. On the other hand, individuality refers to the unique traits and style that distinguish each writer, influencing the tone and content of their work to convey genuineness and effectively engage readers. Striking a balance between personalization and individuality enables writers to create captivating and relatable content that resonates with diverse audiences, while also showcasing their own unique identity.

From writing alone, most of the time, it is possible to understand the inner feelings, characteristics and emotions of the writer. Therefore, it is safe to say every writer has a unique method of writing, which establishes their writing style. While creativity is a big factor in enhancing the quality of writing, some programs tend to develop rigid rules, restricting the level of individuality and personalization. Therefore, critical thinking abilities are further improved if one is encouraged to add a personal touch to one’s writing \cite{krueger2001content-2}. Most notably, for educational purposes, personalization plays a big role in ensuring teachers are giving proper constructive feedback catered to each student \cite{wang2022personalization-3} and students are putting their own thoughts forward rather than blatantly plagiarizing \cite{milano2023large-4}. The recent emergence of AI-assisted tools for writing task completion has observed millions of users employing them for writing purposes. Spellcheckers \cite{wang2021towards-5} were among the first computational writing aids. Further advances include techniques for determining writing style \cite{sterman2020interacting}, ideas for brainstorming support tools \cite{gero2019stylistic,gero2019metaphoria}, and theories of cognitive writing for creating writing frameworks \cite{hui2018introassist}. With all these tools, it may be observed that the writing style of individuals might become similar or lose their individual rhythm, making it monotonous and giving content traits that are suggestive of AI-driven creation and synonymous with automated generation. 

In this work, we use the term "Personal Touch" frequently. This term refers to the unique voice, style, and perspective that an individual writer brings to their work. It encompasses a distinct tone, relatable personal anecdotes or experiences, clear opinions and viewpoints, preferred writing style and word choice, as well as the ability to forge an emotional connection with the reader. The personal touch is what allows a writer's personality and humanity to shine through their words, distinguishing their writing from others and making it feel more authentic, intimate, and engaging.

\section{Personalization in the Age of LLMs}
LLMs show potential for helping users with a range of creative writing activities, including screenplays \cite{mirowski2023co} and short stories \cite{yang2022re3, wordcraft}. LLMs have become useful tools for helping users without a background in computing by responding directly to user commands after being aligned to output content more in accordance with human preferences \cite{ouyang2022training}. Nevertheless, modern LLMs continue to lack a robust fact-checking method, which could result in errors, mainly when writing non-fiction \cite{jiang2020can}. From what appears to be harmless information, they can create detrimental details \cite{gehman2020realtoxicityprompts, perez2202red}, putting minorities in danger of being misrepresented. Sociopolitical and environmental challenges also come up as ethical issues \cite{bender2021dangers}. Although prompt engineering is becoming more popular to handle these issues \cite{bach2022promptsource, liu2023pre}, non-technical users may find it intimidating due to its intricacy. Another concern of LLM usage for writing is prominent in academia is plagiarism \cite{park2017other}, where education bodies must decide how to react: ban it or incorporate it in the teaching and examining systems \cite{milano2023large-4}. Padmakumar and He \cite{padmakumar2023does} conduct an experiment to contrast writing between model-free assisted writing, feedback-tuned writing (InstructGPT), and utilizing a basic model (GPT-3). InstructGPT’s results indicate that it considerably reduces diversity, increasing the similarity of texts written by different writers. This decrease results from the model producing content that is less varied, even when user contributions stay the same. 

\begin{figure*}[t] 
\centering {\includegraphics[width=\textwidth]{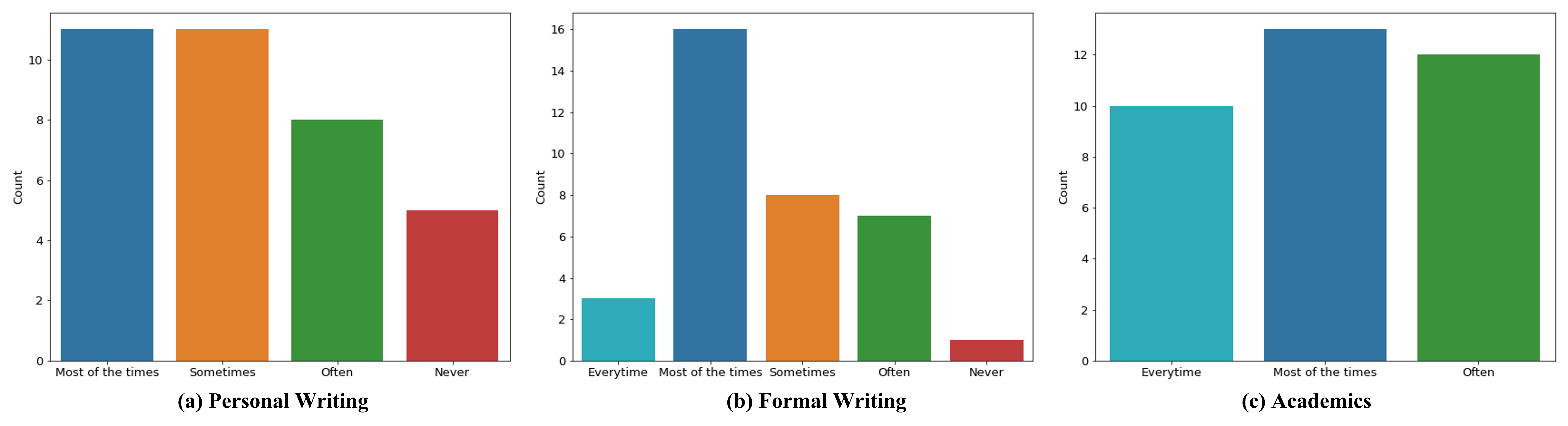}}
\caption{Survey Statistics: Frequency of LLM Assistance in Different Types of Writing}
\label{fig:1_2_3}
\end{figure*}

\begin{figure*}[t] 
\centering {\includegraphics[width=\textwidth]{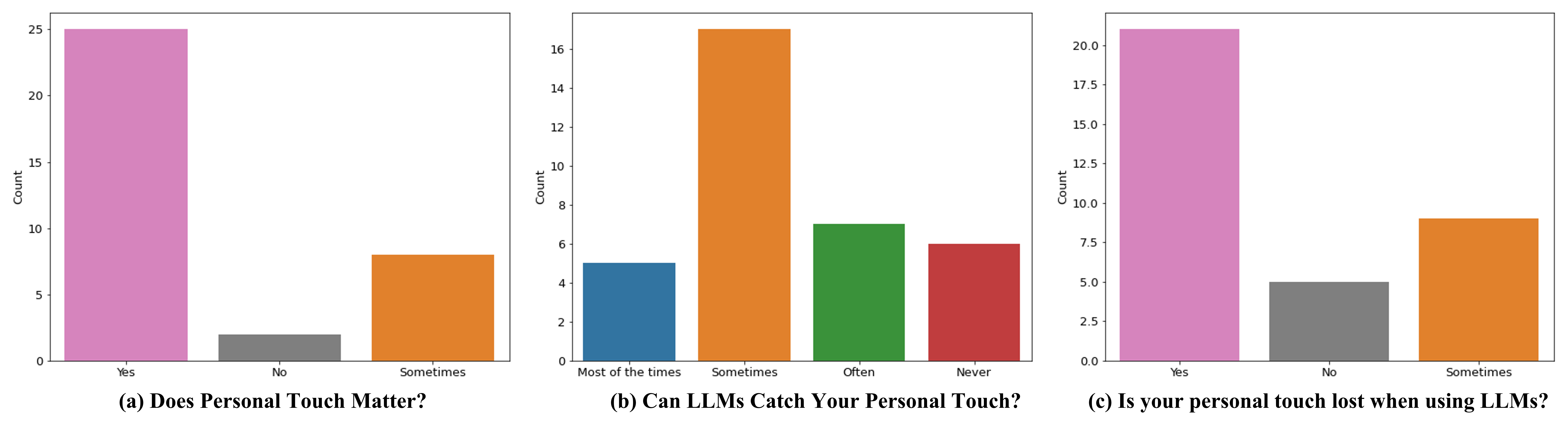}}
\caption{Survey Statistics: View on Personalization and LLMs}
\label{fig:5_6_7}
\end{figure*}

\begin{figure}[t] 
\centering {\includegraphics[scale=0.35]{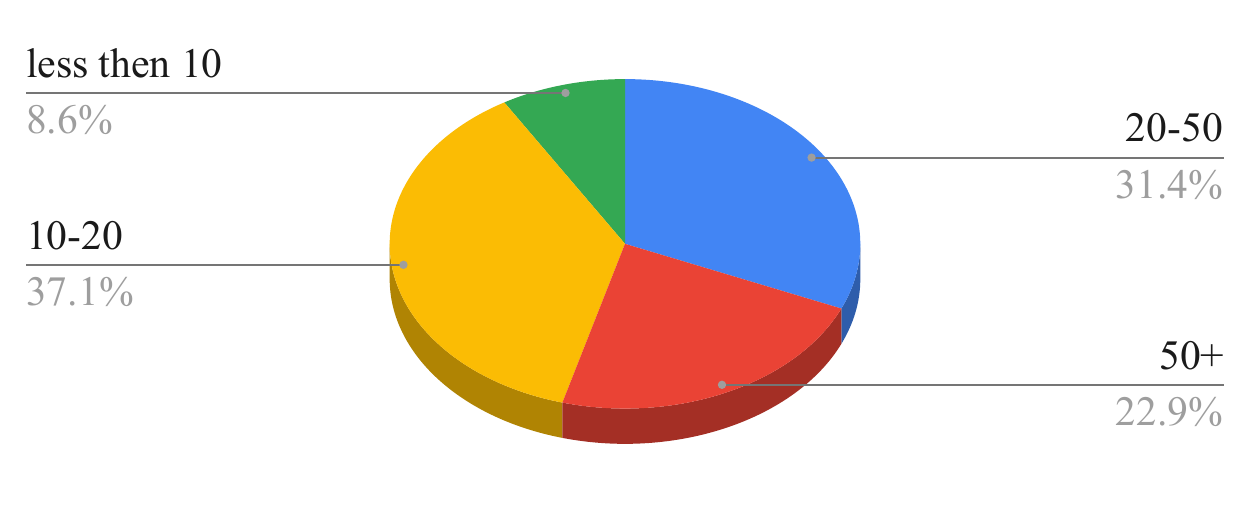}}
\caption{Average prompt length by survey participants}
\label{fig:pl}
\end{figure}

\section{Problems and Perspectives}
\subsection{Survey Information}
To investigate the role of LLMs in content personalization and individuality, we conducted a brief survey with 35 participants. The demographic profile of the participants are available in Appendix  \ref{sec: demographics}. The responses that we received show different perspectives and concerns in this area. Figure \ref{fig:1_2_3} shows the frequency of LLM assistance in three types of writing: personal writing (letters, messages and birthday wishes), formal writing (CVs and applications), and academic writing (reports, assignments and homework).

\subsection{Survey Questions} \label{sec: sur-q}
Below are the survey questions included in the study:
\begin{itemize}
    \item Personal Writing: \textit{How often do you write personal notes (letters, messages, birthday wishes) using ChatGPT/other AI  writing assistants?} \\
    Options: Every time, Most of the time, Often, Some times, Never.
    \item Formal Writing: \textit{How often do you use it for formal writings (job tasks, applications and stuffs)?} \\
    Options: Every time, Most of the time, Often, Some times, Never.
    \item Academic Writing: \textit{How often do you use it for academics (reports, assignments, anything)?} \\
    Options: Every time, Most of the time, Often, Some times, Never.
    \item Average prompt length by survey participants : \textit{Your average prompt length (word count)} (to assess context and modifications) \\
    Options: less than 10 words, 10-20 words, 20-50 words, 50+ words
    \item Can LLMs catch you personal touch? : \textit{Do you think AI assistants can catch your personal touch (your writing style, or feelings)? (As they generate general outputs most of the times)}\\
    Options: Every time, Most of the time, Often, Some times, Never.
    \item Does personal touch matter? \textit{Do you think your personal touch in writing matters in different types of writing?}\\
    Options: Yes, No, sometimes.
    \item Is your personal touch lost when using LLMs? \textit{Do you think your personal touch in writing is being lost while using AI?}\\
    Options: Yes, No, sometimes.
\end{itemize}

\subsection{Survey Statistics and Interpretation}
The survey reveals widespread use of LLMs across various writing contexts. Personal writing sees the least utilization (14.3\% people never use LLMs) due to a desire to maintain original emotional expression. Conversely, LLM usage is nearly ubiquitous in formal writing, with only around 3\% abstaining, and even more prevalent in academic writing, where it's becoming a standard practice. While content personalization remains crucial in all contexts, its significance appears to vary, with academic settings showing a trend towards increased reliance on LLMs (28.6\% use them everytime), hinting at a potential evolution in educational practices.

So, \textbf{\textit{do LLMs negatively affect personalization and individuality?}} Our survey data reveals that a majority (71.4\%) value personal touches in writing (Figure \ref{fig:5_6_7} (a)), and many believe LLMs can capture their personal touch (Figure \ref{fig:5_6_7} (b)). However, despite this, 60\% of respondents feel their personal touch is lost when using LLMs (Figure \ref{fig:5_6_7} (c)), suggesting a discrepancy between perception and reality regarding the impact of LLMs on personalization and individuality. Also, figure \ref{fig:pl} illustrates that the majority of prompts are relatively short in length, suggesting that participants may not have employed sufficient prompt engineering techniques to generate more personalized or engaging content.

So, the answer is yes. Extensive use of LLMs in writing diminishes personalization and individuality by homogenizing content. These AI systems often prioritize efficiency and conformity over unique expression, leading to a loss of the author's distinct voice and style. Furthermore, reliance on LLMs can discourage individuals from cultivating their writing skills and developing their own creative approaches, resulting in a dependency that hampers personal growth and self-expression. Additionally, the standardized nature of LLM-generated content may dilute the emotional resonance and authenticity that characterize truly personalized writing, detracting from its impact on readers.

\subsection{Observation}
Respondents recognize the value of personalization in writing and believe LLMs can capture some aspects of personal style, yet many still feel that their unique touch gets lost when using LLMs. This discrepancy highlights potential limitations in current LLMs' ability to faithfully replicate individual expression. One of the reasons for believing LLMs can capture personal touches is similarity in tone or structure. However, participants feel that their personal style is lost because the individuality in their expression or unique writing habits are not accurately reflected, resulting in a sense of impersonal or generic writing for all.
Additionally, participants may feel a decline in their own writing or thinking skills when relying on LLMs, contributing to concerns about losing personal touch. These findings underscore the need for more research and development to address challenges in maintaining personalization in AI writing tools and raise doubts about LLMs' effectiveness in preserving authenticity and uniqueness in personal writing.

\section{Why does losing our individuality matter?}
From our survey, we have discovered that given three modes of writing: personal, academic and formal, they prefer to leverage LLMs to generate content. One interesting finding is that for personal writing, we have the same number of individuals using LLMs sometimes and most of the time. But what we can conclude from our findings is that the majority of users aren’t bothered too much about LLMs not being the greatest agents of personalized writing. As long as informative information is captured, they are satisfied. 

But, lack of personalization with extensive LLM usage can lead to generic, impersonal content that fails to engage readers on a deeper level or resonate with their specific interests and needs \cite{Bekker2024}. Similarly, a lack of individuality can result in homogenized writing that lacks uniqueness and fails to showcase the author's voice and perspective. Over time, this may lead to decreased reader interest, reduced credibility, and a loss of connection between the writer and audience. Therefore, raising awareness about the importance of personalization and individuality in writing is crucial for fostering more meaningful and impactful communication.

\section{Related Works and Discussion}
From our surveys, we have established the need for LLMs to capture personalization and reflect in for the various forms of writing: personal, formal and academic. With most of our survey indicating users want personified writing, we hope LLMs which aides in such personalization are constructed because with LLMs becoming a part of our daily and professional lives, it is imperative these AI-assisted writing tools are not giving us monotonous and generic content. One such collaborative group-AI Brainwriting brainstorming framework was developed by \citet{shaer2024ai}. They included an LLM to improve the group ideation process, and then they assessed the process of idea generation along with the resulting solution space. With creativity being a main priority to be included with LLMs, \citet{xu2024jamplate} propose Jamplate, a digital whiteboard plugin with LLM integration, to simplify the idea generating process which is an an alternate strategy that substitutes chat interfaces with structured templates is put forth. In their preliminary research, they discovered Jamplate helps inexperienced designers think more critically and improve their ideas. With such works existing to tackle the existing issues of LLMs, we believe our work is a step towards the right direction and brings attention to address the lack of creativity and human-touch like feel to automated content writing. Additionally, this is a crucial topic for discussion considering the presence of other widely spoken languages globally such as Chinese, French, German, Hindi, and more. Given the widespread international use of LLMs, this issue requires prompt attention and resolution.

\textbf{Limitations.}
One drawback of this study is the limited sample size of only 35 participants. Enhancing the number of participants could enhance the robustness and significance of the research outcomes. Also, the survey questions can be more effectively designed. We hope to work on these issues in future.

\section{Conclusion}
This study investigates the different obstacles and perspectives of writing with personalization and individuality in the age of LLMs. According to our survey, we are currently heavily reliant on LLMs for writing any kind of content, from personal to format to academic; and this heavy use of LLMs is harming our individuality, but we are not too concerned about it. This study connects our survey with literature to understand individual perspectives and contribute to the development of personalised writing with LLMs.
Below is a concise summary of both the data-supported findings regarding participants' motives and some speculative insights:
\begin{itemize}
    \item We investigate whether writers are concerned about adding personalized content to their writings and if LLMs can capture it. This research suggests a discrepancy between the perceived and actual impact of LLMs on personalization and individuality in writing. We find out most writers have claimed that personal touch is important in writing, but LLMs fail to capture their persona and nature for content writing, indicating that extensive reliance on LLMs homogenizes content and diminishes individual expressions.
    \item We also discover that writers are not too burdened by LLMs' lack of ability to capture their individuality and personal touch as long as the main gist is identified, which is very concerning.
\end{itemize}
In observation, we emphasize the need for features or modes in current LLMs that will help obtain the personal touch in writing to embrace an individual's uniqueness. Future discussions and research on this area can focus on this aspect.

\section*{Author Contributions}
ATW conceived the core idea, developed the methodology, designed the experiments and the data collection, performed some experiments, conducted all the analyses, prepared visualizations, wrote and edited core parts of the paper, and led the whole project. 
RI helped in background study, formal analysis, initial draft writing and editing.
MRI collected the data and was involved in editing.


\bibliographystyle{ACM-Reference-Format}
\bibliography{our_work}

\appendix

\section{Survey Participant Demographics} \label{sec: demographics}
The demographic data of survey participants presents a demographic consisting primarily of university students aged 18-24 (94.3\%), with a smaller proportion (5.7\%) in the 24-30 age group. The gender distribution is skewed towards males (65.7\%), with females making up 34.3\% of the group. Notably, the overwhelming majority (88.6\%) come from a Science background, while a small percentage (8.6\%) are from Arts backgrounds . Significantly, most individuals in this demographic rated their technology awareness as high, scoring between 3 and 5 on a scale where 5 represents a "tech nerd" and 1 indicates being unaware of technology. This suggests a cohort of tech-savvy individuals, likely pursuing STEM fields in their university studies.








\end{document}